\documentclass[12pt,preprint]{aastex}
\singlespace

\shortauthors{Phillips \& Kogut}
\shorttitle{WMAP Constraints on Global Topology}

\begin{document}

\title{Constraints On The Topology Of The Universe From The 
WMAP First-Year Sky Maps}

\author{N. G. Phillips\altaffilmark{1},
A. Kogut\altaffilmark{2} }

\altaffiltext{1}{SSAI, Goddard Space Flight Center, Greenbelt, MD 20771}
\altaffiltext{2}{Code 685, Goddard Space Flight Center, Greenbelt, MD 20771}

\email{Nicholas.G.Phillips.1@gsfc.nasa.gov}

\begin{abstract}
We compute the covariance expected
between the spherical harmonic coefficients $a_{\ell m}$
of the cosmic microwave temperature anisotropy
if the universe had a compact topology.
For fundamental cell size smaller
than the distance to the decoupling surface,
off-diagonal components carry more information
than the diagonal components
(the power spectrum).
We use a maximum likelihood analysis
to compare the Wilkinson Microwave Anisotropy Probe
first-year data to models
with a cubic topology.
The data are compatible with
finite flat topologies
with fundamental domain
$L > 1.2$ times the distance to the
decoupling surface at 95\% confidence.
The WMAP data show reduced power at
the quadrupole and octopole,
but do not show the correlations expected
for a compact topology
and are indistinguishable from infinite models.
\end{abstract}



\section{INTRODUCTION}
The simplest model for the universe
is a spatially homogeneous, isotropic spacetime
with a Euclidian (flat) geometry.
This simple model
is consistent with observations,
but leaves unaddressed the question of topology
or the connectedness of spacetime.
Schwartzschild 
\citep{schwartzschild:1900}
first noted the possiblitity of non-trivial
topology for the Universe even before
Einstein's discovery of his field equations.
Almost immediately after Einstein's discovery,
de Sitter \citep{desitter:1917} pointed out the
the field equations did not constrain the topology.
Since general relativity provides no theoretical guidance,
we turn to observations
to constrain topology.

Observational tests of topology
all rely on multiple imaging of distant objects.
If the universe is multiply-connected
with cell size smaller than the distance to some object,
photons from that object can reach the observer
via multiple paths.
Simply searching the sky for multiply-imaged point sources, e.g. quasars, 
is problematic: since the travel time to each image is different, 
each image shows the same object at a different time.  
If source evolution is important,
the multiple images may no longer be recognizable as such.
The ideal source for topological tests
would fill the whole sky
with a pattern centered on the observer
and emitted at a single time.
The cosmic microwave background (CMB) is an excellent
approximation to this ideal source.

A number of authors have used the CMB
to constrain the topology of the universe.
These tests fall into two general categories.
A compact topology
can not support spatial structure
with wavelength longer than the cell size.
The CMB temperature anisotropy
will thus be suppressed
on angular scales larger than the (projected) cell size.
The first category tests use the CMB power spectrum
(or its Legendre transform, the 2-point correlation function)
to test for non-trivial topology.

The CMB in fact
shows significantly less power
in the quadrupole and octopole
than would be expected for a model
based on higher-order moments.
The discrepancy was first detected
by the Cosmic Background Explorer
\citep{bennett/etal:1996}
and verified at much high signal to noise ratio
by the 
Wilkinson Microwave Anisotropy Probe (WMAP)
\citep{bennett/etal:2003}.
Figure \ref{fig1-powerspectrum}
shows the angular power spectrum
of the WMAP first-year data
compared to the best-fit $\Lambda$CDM model
\citep{spergel/etal:2003}.
Models with compact topology $L \sim 1$
provide a good match
to the observed power spectrum,
motivating tests of finite-universe models
\citep{oliveira/smoot:1995,
luminet/etal:2003}.

The suppression of power on large angular scales
is a necessary but not sufficient condition
for the existence of a compact topology.
The power spectrum is rotationally invariant,
averaging over any phase information in the pattern of CMB anisotropy.
Such phase information must exist for compact topologies,
and forms the basis for a second class of tests.
A ``circles on the sky'' search
\citep{cornish/etal:1998}
provides a more stringent test for compact topologies.
The CMB decoupling surface
is a sphere centered on the observer.
If the cell size is smaller than the distance to the decoupling surface,
the multiple images of this sphere induced by a compact topology
will intersect
to produce patterns that match along circles.
Such circles are not observed,
limiting the cell size $L > 1.7$
for a wide class of models
\citep{cornish/etal:2003}.

Additional tests are possible.
Compact topologies will not produce circles on the sky
if the cell size is larger
than the distance to the source,
since the resulting images will not intersect.
Compact topologies with $L > 2$
may still be distinguished 
using phase information.
In this paper,
we describe the correlations imposed on the microwave background
by the topology.
We use this formalism to compare the WMAP first-year data
to a model with cubic topology
and derive constraints on the cell size $L$.

\section{COVARIANCE OF SPHERICAL HARMONIC COEFFICIENTS}

On large scales, the CMB temperature anisotropy 
is given by
\begin{equation}
\Delta T(\hat{\bf x}) = -\frac{1}{3} \Phi( {\Delta\tau}\, \hat{\bf x}),
\end{equation}
where $\Phi({\bf x})$ is the gravitational potential and ${\Delta\tau}$ is the
radius of the decoupling surface. 
The potential has the harmonic expansion
\begin{equation}
\Phi({\bf x}) = \int{\!d\mu({\bf k})\;} \Phi_{{\bf k}}\; e^{-i{\bf k}\cdot{\bf x}}
\label{eq-Phi-expand}
\end{equation}
where $\;{\!d\mu({\bf k})\;}$ reflects the density of states,
determined by the topology. 
For an infinite universe,
${\bf k}$ is continuous,
while for a compact topology
only discrete values
\begin{equation}
{\bf k} = \frac{2 \pi }{L} {\bf n},\quad
{\bf n} = \left(n_x, n_y, n_z \right)
\label{discrete_k_eq}
\end{equation}
are allowed, where
$n_x$,$n_y$,and $n_z$ are integers
and $L$ is the cell size
in units of the conformal time to the decoupling surface
\citep{zeldovich:1973,
fang/mo:1987,
sokolov:1993}.
The cutoff in the discrete spectrum,
$|{\bf k}| > \frac{2 \pi}{L}$,
suppresses power on large angular scales.

In inflationary cosmologies,
the gravitational potential's $\Phi_{{\bf k}}$
are random Gaussian variables with zero mean and the covariance
\begin{equation}
\left< \Phi_{{\bf k}} \Phi^*_{{\bf k}'} \right> =
    \frac{2\pi^2}{k^3}\delta^3({\bf k}-{\bf k}')\, {\cal P}(k).
\label{eq-Phi-covar}
\end{equation}
We expand the corresponding temperature fluctuations
\begin{equation}
\Delta T(\hat{\bf x}) = \sum_{\ell m} a_{\ell m} Y_{\ell m}(\hat{\bf x})
\label{ylm_eq}
\end{equation}
where
\begin{equation}
a_{\ell m} = -\frac{(-i)^\ell 4\pi}{3} \int{\!d\mu({\bf k})\;}
             \Phi_{{\bf k}}\; j_\ell(k{\Delta\tau})\; Y_{\ell m}({\hat {\bf k}})
\label{eq-alm}
\end{equation}
and $ j_\ell(k{\Delta\tau}) $
is a Bessel function of order $\ell$.
The $a_{\ell m}$'s have zero mean.
Using Eqn (\ref{eq-Phi-covar}), we find their covariance
\begin{eqnarray}
{\bf M}^L_{\ell m,\ell'm'} &=&
\left< a_{\ell m} a^*_{\ell'm'} \right> \\
&=&
 \frac{(-i)^\ell i^{\ell'} 32\pi^4}{9}
      \int{\!d\mu({\bf k})\;}
         j_\ell(k{\Delta\tau})j_{\ell'}(k{\Delta\tau}) \;
         \frac{{\cal P}(k)}{k^3} \;
         Y_{\ell m}({\hat {\bf k}}) Y^*_{\ell'm'}({\hat {\bf k}}),
\label{eq-alm-covar-form1}
\end{eqnarray}
which is not necessarily diagonal.

In the limit of a flat open Universe,
${\!d\mu({\bf k})\;} \rightarrow k^2 dk\,d{\hat {\bf k}}$ and
the orthonormality of the spherical harmonics yields
\begin{equation}
\lim_{L\rightarrow\infty} {\bf M}^L_{\ell m,\ell'm'} =
    \frac{ 32\pi^4}{9}
	\delta_{\ell\ell'}\delta_{mm'}
    \int_0^\infty\!dk \; j_\ell(k{\Delta\tau})^2 \; \frac{{\cal P}(k)}{k},
\end{equation}
For a compact topology,
the integral over the continuous variables $\bf k$
becomes a sum over discrete ${\bf k}_n = (2\pi/L){\bf n}$
with ${\bf n} = (n_x,n_y,n_z)$ a triplet of integers.
Just as we break the continuous integration ${\!d\mu({\bf k})\;}$ into
magnitude and angular parts, we do the same for the discrete case:
${\!d\mu({\bf k})\;} = \sum_{n\in{\cal N}} \sum_{\{|{\bf n}| = n\}}$ where
${\cal N}$ is the set of all possible magnitudes of the integer
triplets $\bf n$ and $\{|{\bf n}| = n\}$ are all $\bf n$ with magnitude
$n$. Thus $\sum_{n\in{\cal N}}$ is the sum over magnitude and
$\sum_{\{|{\bf n}| = n\}}$ is, for each magnitude, the angular sum.
For example, the first value in $\cal N$ is $n=1$ with the
corresponding $\{|{\bf n}| = 1\}$ containing the six vectors
$(\pm 1,0,0)$, $(0,\pm 1,0)$ and $(0,0,\pm 1)$. The next value
in $\cal N$ is $n = \sqrt{2}$ and $\{|{\bf n}| = \sqrt{2}\}$ contains
12 vectors, each with $\pm 1$ in two places and $0$ in the third.
Writing the measure ${\!d\mu({\bf k})\;}$ in this form, for compact topologies
we have
\begin{equation}
{\bf M}^L_{\ell m,\ell'm'} =
    \frac{32\pi^4}{9}
	\sum_{n\in{\cal N}}
            \; j_\ell(k_n {\Delta\tau})  j_{\ell'}(k_n {\Delta\tau})
            \; \frac{{\cal P}(k)}{k^3}
			\; A^L_{\ell m,\ell'm'}(n)
\label{eq-alm-covar-jl}
\end{equation}
where
\begin{equation}
A^L_{\ell m,\ell'm'}(n) = (-i)^\ell i^{\ell'} \sum_{\{|{\bf n}| = n\}}
                           Y_{\ell m}({\hat{\bf n}}) Y^*_{\ell'm'}({\hat{\bf n}})
\label{split_eq}
\end{equation}
The momentum $k_n$ is related to $n$
via $k_n = \frac{2\pi}{L}|{\bf n}|$.

%
There is a great computational advantage for making this split in the sum.
All the cosmology is contained in the magnitude sum while
the topology is reflected
in the sum for $A_{\ell m,\ell'm'}(n)$,
which requires the lion's share of CPU time.
When numerically evaluating Eqn (\ref{eq-alm-covar}),
we compute and store all the
required matrices $A_{\ell m,\ell'm'}(n)$,
which only depend on the relative ratios of the
fundamental domain sizes $L_x$, $L_y$ and $L_z$.
For this work, we have assumed
they are all the same and equal to $L$.
As we vary the rest of the model parameters,
including the topology scale $L$,
we need only evaluate the factors in the first sum
and use the stored matrices $A_{\ell m,\ell'm'}(n)$ to complete the sum.

Thus far, we have ignored contributions
from the time evolution of the gravitational potential
(the integrated Sachs-Wolfe effect)
on large scales
and from the Boltzmann physics of the coupled photon-baryon fluid.
on small scales.
Both effects may readily be included.
The ``line-of-sight'' approach \citep{seljak/zaldarriaga:1996}
to computing the anisotropies for a flat open Universe first calculates
the (scalar) response function $\Delta^{(S)}_{T\ell}(k,\Delta\tau)$ and then gives the
(diagonal) covariance
\begin{equation}
{\bf M}_{\ell m,\ell'm'}^{\rm CMBFAST} =
    (4\pi)^2
	\delta_{\ell\ell'}\delta_{mm'}
    \int_0^\infty\!dk \; |\Delta^{(S)}_{T\ell}(k,\Delta\tau) |^2
                      \; \frac{{\cal P}(k)}{k}
\label{eq-standard-alm-covar}
\end{equation}
This suggests we can include the ISW and acoustic effects of the plasma motion
by writing covariance (\ref{eq-alm-covar-form1}) as
\begin{equation}
{\bf M}^L_{\ell m,\ell'm'} =
	(4\pi)^2 \sum_{n\in{\cal N}}
	        \,\Delta^{(S)}_{T\ell}(k_n,\Delta\tau)\Delta^{(S)}_{T\ell'}(k_n,\Delta\tau)
            \; \frac{{\cal P}(k)}{k^3}
			\; A^L_{\ell m,\ell'm'}(n)
\label{eq-alm-covar}
\end{equation}
with $A^L_{\ell m,\ell'm'}(n)$ given by Eqn (\ref{split_eq}).
We obtain $\Delta^{(S)}_{T\ell}(k,\Delta\tau)$
from CMBFAST \citep{seljak/zaldarriaga:1996}, by way of the program CMBEASY \citep{doran:2003},
along with the associated value of ${\Delta\tau}$.

Equation \ref{eq-alm-covar} gives the covariance of the
temperature coefficients $ a_{\ell m}$
as a function of cosmological parameters
and global topology.
Figure \ref{fig2-off-diag-power} compares
the diagonal elements
to the off-diagonal elements
for the case of a cubic topology.
The total power on the diagonal,
$\sum_{\ell m}|{\bf M}^L_{\ell m \ell m}|$, 
reflects the content of power spectrum $C^L_{\ell}$ 
and is rotationally invariant. 
The off-diagonal power,
$\sum_{(\ell m)\ne(\ell' m')}|{\bf M}^L_{\ell m \ell' m'}|$,
measures the correlations between the different angular scales 
and is generated by the global properties of the topology. 
For cell size $L$ less than twice the distance to the
decoupling surface,
the off-diagonal elements dominate.
Analyses based solely on the power spectrum
thus ignore the main information content of the map.
The off-diagonal correlations persist at larger cell size,
but become increasingly less important.
In contrast to the circles on the sky test,
the off-diagonal correlations
smoothly decrease past $L > 2$.

\subsection{Sky Map Generation }
The covariance matrix ${\bf M}^L_{\ell m,\ell'm'}$
for the $a_{\ell m}$ coefficients
fully describes a cosmological model and topology.
Since the $a_{\ell m}$'s are
still Gaussian variables, albeit now correlated, all higher moments are given
in terms of their covariance.
Equation \ref{split_eq} thus allows the rapid generation
of simulated sky maps for a given topology.
For a given topology scale $L$ and cosmology 
$\Delta^{(S)}_{T\ell}(k)$,
we first compute the Cholesky decomposition 
${\bf L}_{\ell m \ell'm'}$
of the $a_{\ell m}$ covariance matrix:
${\bf M}_{\ell m \ell'm'} 
= {\bf L}_{\ell m \ell''m''}\;{\bf L}_{\ell'm'\;\ell''m''}$.
Then if $x_{\ell m}$ are a set of 
uncorrelated unit variance Gaussian variables, we set
$a_{\ell m} = \sum_{\ell' m'}{\bf L}_{\ell m \ell'm'}\;x_{\ell'm'}$ 
to obtain the sky map  
$\Delta T({\hat{\bf x}}) = \sum_{\ell m} a_{\ell m} Y_{\ell m}({\hat{\bf x}})$.
A ``circles in the sky'' test
\citep{cornish/etal:1998} 
run on the resulting sky maps
verifies that the algorithm correctly reproduces all features
for compact topologies.

The only remaining free parameter in our model is the overall amplitude of the fluctuations. 
We fix this by comparing the power spectrum for the compact model
to that for an infinite Universe with the
same response function $\Delta^{(S)}_{T\ell}(k)$. 
The CMB power spectrum is given by
\begin{equation}
C_\ell = \frac{1}{2\ell+1} \sum_m \left< |a_{\ell m}|^2 \right>
        = \frac{1}{2\ell+1} \sum_m M^L_{\ell m\ell m}.
\label{power_spec_def}
\end{equation}
The covariance of the power spectrum becomes
\begin{equation}
{\bf M}_{\ell\ell'}^{L,C_\ell} = \frac{2}{(2\ell+1)(2\ell'+1)}
                               \sum_{m=-\ell}^\ell \;\sum_{m'=-\ell'}^{\ell'}
							   \left|{\bf M}^L_{\ell m\ell' m'}\right|^2,
\label{eq-Cl-covar}
\end{equation}
We normalize the covariance matrices ${\bf M}^L_{\ell m\ell' m'}$
using the amplitude that minimizes
$\sum_{\ell_{lower}}^{\ell_{max}}
\left( C^L_\ell - C^{\rm CMBFAST}_\ell\right)^2$
where we use $\ell_{lower}=20$, as this is a small enough scale to be beyond
where we expect to find interesting topological effects.
The simulated maps correctly show the rise to the first Doppler peak.
Since topological effects are most apparent on large scales,
we take $\ell_{max}=30$ for all analyses in this paper.

\section{LIKELIHOOD ANALYSIS}

We compare data from the WMAP first-year sky maps
to a set of models with a cubic fundamental domains
described by the cell size $L$
in units of the conformal time to the decoupling surface.
We specialize to the case of a cubic topology,
identifying opposing faces of the unit cell
without twists or rotation,
so that the topology is fully specified
by the cell size $L$.
For each value $L$ we compute the likelihood
$ \log{\cal L} =
  -\frac{1}{2}\left( \chi^2 + \log\det{\bf M} \right) $
where
\begin{equation}
\chi^2 = \Sigma_{\ell \ell' m m'}
    ~ a_{\ell m}
    ~ ({\bf M}^{-1})_{\ell m\ell' m'}
    ~ a_{\ell' m'}
\label{chisq_def}
\end{equation}
and
${\bf M}_{\ell m\ell' m'}$
is given by Equation \ref{eq-alm-covar}.

We use temperature coefficients
$ a_{\ell m} $
derived from the internal linear combination (ILC) map
from the WMAP first-year data release
\citep{bennett/etal:2003b}.
This map reduces foreground emission
at the cost of a complicated window function
and instrument noise on angular scales $\theta < 2\arcdeg$.
Topology is important only at much larger angular scales.
We limit the likelihood calculation 
to $2 < \ell < 30$.
On these scales,
the dominant uncertainty is cosmic variance;
effects from the instrument noise and beam profiles are negligible.

The ILC map reduces foreground emission but does not eliminate it completely.
We impose a cut in Galactic latitude $|b| > 5\arcdeg$
and compute the $ a_{\ell m} $
using unit weight for all pixels outside the cut.
Our results are stable as the cut is varied
from 2\fdg5 to 15\arcdeg.

The likelihood analysis
includes off-diagonal correlations
between different $ a_{\ell m} $,
and is not rotationally invariant.
Equation \ref{eq-alm-covar} was derived for the case
when the faces of the fundamental domain
align with the data coordinate system.
We must thus consider different possible orientations
between the data and the unit cell of the topology.
It does not matter whether we rotate the model or the data,
so we rotate the data.
In terms of the Euler angles $\vec\xi = (\phi,\theta,\psi)$, we use
the rotation matrix
\begin{equation}
{\bf R} =
{\bf R}_{\ell m \ell' m'}(\vec\xi) =
   \delta_{\ell \ell'}\; e^{-i\phi m'}\, d_{mm'}^\ell(\theta) \, e^{-i \psi m},
\label{euler_eq}
\end{equation}
where $d_{mm'}^\ell(\theta)$ are the Wigner rotation functions.
We apply the rotation 
on a grid of TBD possible orientations $\vec\xi$
over the range
$0\le \phi,\theta,\psi \le \pi/2$. 
$\phi$ and $\theta$ are uniformly distributed on the sphere
while the azimuthal rotation $\psi$ is uniformly distributed over its range.
Since the fundamental domain for the topology is cubic, 
the noted range covers all possible orientations.

For a fixed cosmological model, 
the likelihood becomes a function of four parameters:
\begin{equation}
\log{\cal L}\left({ {\bf a}}|L,\vec\xi\right) =
 -\frac{1}{2}\left(
        ({\bf R}\,{ {\bf a}})^T
		         \cdot ({\bf M})^{-1} \cdot
        ({\bf R}\,{ {\bf a}})
 + \log\det{\bf M} \right),
\label{eq-alm-likelihood}
\end{equation}
one parameter specifying the size of the unit cell
and three parameters specifying the orientation,
where for clarity we have suppressed
the $\ell m$ subscripts.
Even in the absence of  topological effects,
chance alignments between the random $a_{\ell m}$
can create non-zero correlations in a single realization.
If the relative orientation of the data and model
is held constant,
the likelihood will approximate
a normal distribution
over an ensemble of CMB maps.
Selecting the maximum likelihood
over many possible orientations
of a {\it single} CMB map, however,
selects the lowest $\chi^2$ at each cell size $L$.
If the data do not represent a model with compact topology,
maximizing the likelihood over the (now nuisance) 
rotational parameters $\vec\xi$
will select from the tail of the $\chi^2$ distribution
(Eq. \ref{chisq_def}),
leading to a biased estimate for $L$.
The bias is unimportant
for large $L$ 
where the scale is too large for the topology to break global isotropy,
but becomes significant for $L \sim 2$
where the off-diagonal terms begin to dominate
(Fig. \ref{fig2-off-diag-power}).

We quantify the effect of maximizing the likelihood over orientation
using Monte Carlo simulations.
We generate 1000 realizations of compact topologies
at 51 different values of $L$
on a grid from 
$L=0.8\Delta\tau$ to $4.0\Delta\tau$
uniformly spaced in $1/L$.
For simplicity, 
we work directly with the $a_{\ell' m'}$ coefficients
to avoid the intermediate steps 
of generating sky maps,
masking pixels near the Galactic plane,
and computing the $a_{\ell' m'}$
from the un-masked pixels.
We instead impose the Galactic cut
using the matrix
\begin{equation}
{\bf P} = {\bf P}_{\ell m \ell' m'}(b) =
\left(1+(-1)^{\ell+\ell'+m+m'}\right)\int_{\cos(\pi/2-b)}^1
 P_\ell^m(y) P_{\ell'}^{m'}(y)dy
\label{bcut_matrix_eq}
\end{equation}
which removes all power from the azimuthally symmetric region 
$|\theta - \pi/2| \le b$,
where $P_\ell^m(y)$ are the associated Legendre functions. 
Each realization
thus generates a set of correlated $a_{\ell' m'}$
\begin{equation}
{\bf a}^L = {\bf P} {\bf R} {\bf L}^L {\bf x}
\label{monte_alm_def}
\end{equation}
where 
${\bf L}^L$ is the Cholesky decomposition of the ${a_{\ell m}}$ covariance matrix
(Eq. \ref{eq-alm-covar})
and $\bf x$ is a vector
of zero mean, unit variance Gaussian random variables.
We select a grid uniform in $1/L$
because the topology effects the covariance matrix
${\bf M}$ as $k_n = \frac{2\pi}{L}|{\bf n}| \propto 1/L$.
There is also the advantage of placing the infinite case 
at a finite distance from the region of interest, $L \sim 1$.
A model with $L=3{\Delta\tau}$ is nearly indistinguishable from $L=\infty$.
We vary the topology scale $L$
while keeping the ``background'' cosmology fixed.
Since we only consider large angular scales,
the results are insensitive to the cosmological parameters.
We run the $k_n$ integration in Eqn. (\ref{eq-alm-covar})
for $n \le 90$, which is sufficient for convergence.

Figure \ref{fig3-maxprob} shows the likelihood of the
WMAP first-year data
as a function of domain size $L$.
When maximized over orientation,
the likelihood peaks sharply at $L \sim 2$.
We test for the significance of this peak
by comparing the WMAP results
to 1000 Monte Carlo simulations
of an infinite flat universe.
The simulations also peak at $L = 2$,
demonstrating the bias incurred
when maximizing over orientation.
The WMAP data fall near the mean
of the simulations,
suggesting that the data
are consistent with an infinite universe.
In this case, the orientation $\vec\xi$
becomes a nuisance parameter.
We may then marginalize over the nuisance parameter
by averaging the likelihood over all orientations
(as opposed to selecting the best orientation).
The marginalized likelihood
shows a plateau for $L > 2.1$
with a sharp drop at smaller cell size.
Note that for $L > 3$ the orientation ceases to be important
as the maximum likelihood asymptotically approaches the marginalized likelihood.

Figure \ref{fig3-maxprob} shows that the WMAP likelihood
falls near the mean of Monte Carlo simulations
drawn from a parent population with an infinite topology.
We use additional Monte Carlo simulations
to set upper limits to the allowed size $L$ of the fundamental domain.
We repeat the maximum likelihood analysis
for 19000 Monte Carlo simulations,
1000 at each of 19 values for $L$ 
ranging from 1.01 to 3.7
uniformly spaced in $1/L$.
For each simulation, we get a ``best fit'' topology scale $L_{out}$, 
the value that maximizes ${\cal L}({{\bf a}^{L_{in}}}|L,\vec\xi)$
(Eq. \ref{eq-alm-likelihood}).
We then tabulate the probability 
${\cal P}(L_{out}|L_{in})$
for each input realization with the ``true'' topology scale $L_{in}$ 
to produce the best-fit output scale $L_{out}$. 
We invert this relationship to
derive the probability 
for a given best-fit output $L_{out}$
to be drawn from a parent population with topology scale $L_{in}$:
\begin{equation}
{\cal P}(L_{in}|L_{out}) = \frac{{\cal P}(L_{out}|L_{in})}
       {\int_0^\infty{\cal P}(L_{out}|L_{in})d(L_{in})},
\label{eq-prob}
\end{equation}
where the factor $d(L_{in})$ 
explicitly accounts for the simulations' uniform distribution in $1/L$.

Figure \ref{fig4-probcontours} shows both distributions. 
For $L_{in}<2$, $L_{out}\simeq L_{in}$ and our likelihood
analysis successfully identifies the topology scale. 
On these scales,
the off diagonal components of ${\bf M}_L$ become dominant
(Fig. \ref{fig2-off-diag-power});
the correlations between the different ${a_{\ell m}}$ values
are important and the likelihood
function strongly discriminates models.
A minor identification
$L_{out}\simeq \sqrt{2}\,L_{in}$
is also apparent.
This aliasing of scale is typical and in this case
corresponds to the ratio in sizes between a circle that
just circumscribes a square and the circle that is just contained
in a square.
Such aliasing of topology scales is expected
and is also seen during a circles-in-the-sky analysis.

For $L_{in}>2$, maximizing over orientation
produces output $L_{out} \approx 2.2$
independent of the actual value of $L_{in}>2$.
The correlations between the $a_{\ell m}$
induced by topology
are weak on scales $L > 2$
compared to chance alignments.
Individual realizations with $L > 2$
will thus appear to have a slight preference for orientation, 
although the parent population is nearly indistinguishable
from the infinite model.
We account for this bias using Eq. \ref{eq-prob}.
An observed best-fit value $L_{out}\sim 2$
does {\it not} mean the parent population
must necessarily have $L_{in}  = 2$,
but rather that the parent population
has nearly uniform probability to represent
any scale $L_{in}\ge 2$.

The maximum likelihood for the WMAP data
occurs for
$L^{WMAP}_{out}=2.1$.
From Eqn. (\ref{eq-prob}),
the cumulative probability is
$P(L_{in} < L_{WMAP,in}) = \int_0^{L_{in}}
 {\cal P}(L'_{in}|L_{WMAP,out})d(L'_{in})$.
We obtain the 68\% confidence limit
that the topology scale is greater than
2.1 times the distance to the decoupling surface
and 95\% confidence it is greater than 1.2 times the distance.
We place a 68\% (95\%) confidence that the topology scale
is greater than 29 Gpc (17 Gpc).

Figure \ref{fig4-probcontours}a
shows that a certain number of realizations with $L < 2$
``scatter''into the region
$L_{out} \sim 2$
where the WMAP data show maximum likelihood.
We test the null hypothesis
(that the universe is infinite)
by taking all simulations
with $L_{out} > 1.96$
(the horizontal band in Fig \ref{fig4-probcontours}a)
and computing the cumulative
probability for these simulations only.
Figure \ref{fig5-wmapprob} shows the resulting curve.
The cumulative probability for the WMAP data,
computed using all simulations,
has identical confidence intervals
as the probability derived using
only those simulations
whose likelihood peak
occurs at $L_{out} > 1.96$.
We therefore accept the null hypothesis
to conclude that the WMAP data
are consistent with an infinite universe.

\section{Discussion}

A compact topology imposes a specific pattern of correlations
$\left<{a_{\ell m}}\, a_{\ell' m'}\right>$
between the spherical harmonic expansion of the CMB temperature.
We compute the expected correlations for the
simplest non-trivial topology, the cubic torus,
and compare a range of cell size $L$
to the WMAP first-year data
using a maximum-likelihood algorithm.
The covariance matrix explicitly includes
the contribution from
the integrated Sachs-Wolfe effect on large angular scales
and the acoustic peaks at small scales.
We separate the covariance
into a piece dependent on the topology
and a piece dependent on the cosmology.
Although the transfer functions $\Delta^{(S)}_{T\ell}(k)$
for the cosmology
assume isotropy in $k$-space,
which is no longer exact for compact models,
the effect is predominantly in the cosmology
with negligible effect on the topology.

\begin{figure}
\plotone{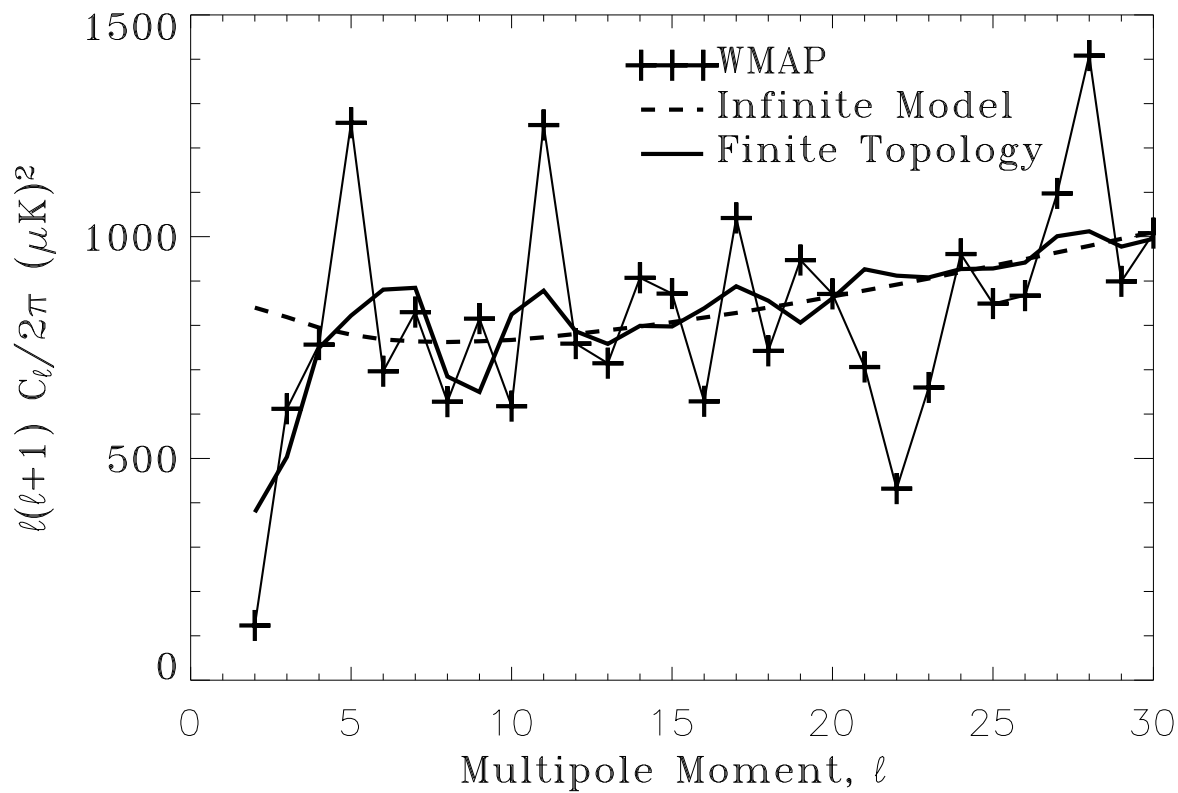}
\caption{Power spectrum of the WMAP first-year data,
compared to a model for a flat universe with an finite topology. 
The topology has a cubic fundamental domain with a side length 
$L = 1.1 ~\times$  the distance to the decoupling surface,
the best-fit value if only the power spectrum is considered.
The dash-dotted line shows the best-fit power spectrum for an open flat Universe.
The lowest three multipoles are suppressed for the finite topology model,
since the universe in such a model is too small
to support power at such large scales.}
\label{fig1-powerspectrum}
\end{figure}

\begin{figure}
\plotone{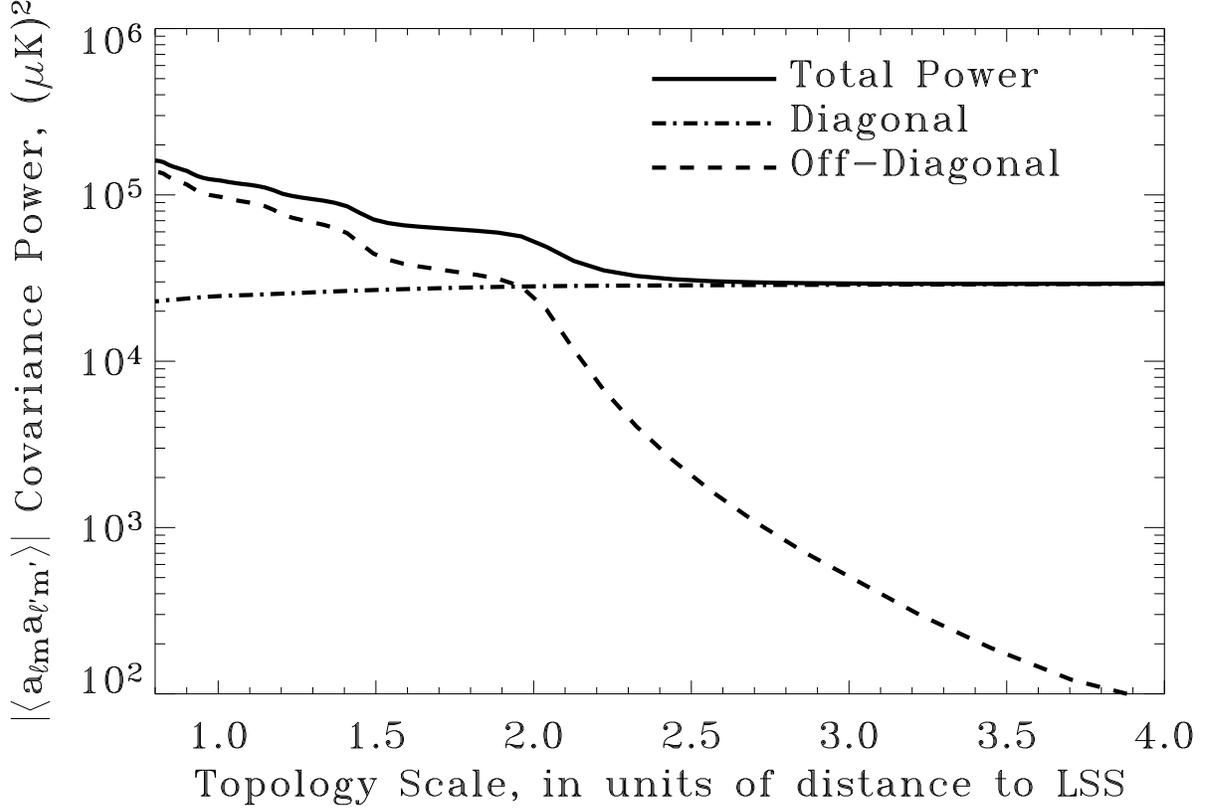}
\caption{Comparison of diagonal vs off-diagonal power
in the spherical expansion coefficient correlation matrix
$|{\bf M}^L_{\ell m,\ell'm'}| = |\left< a_{\ell m} a^*_{\ell'm'} \right>|$ 
as a function topology scale $L$. 
The power measures the relative importance of terms.
The total power in the diagonal components,
$\sum_{\ell m}|{\bf M}^L_{\ell m \ell m}|$, 
measures the information considered by the power spectrum
and is flat with only a small decrease at small $L$ 
due to suppression of power at the largest scales for small topologies.
The off-diagonal power, 
$\sum_{(\ell m)\ne(\ell' m')}|{\bf M}^L_{\ell m \ell' m'}|$,
arises from the global structure present for finite topologies. 
This varies from almost negligible at the largest topology scales 
to being the dominant contribution for small topologies. 
An analysis looking for topology via a spherical expansion 
must consider the correlations between different expansion coefficients.
}
\label{fig2-off-diag-power}
\end{figure}

\begin{figure}
\plotone{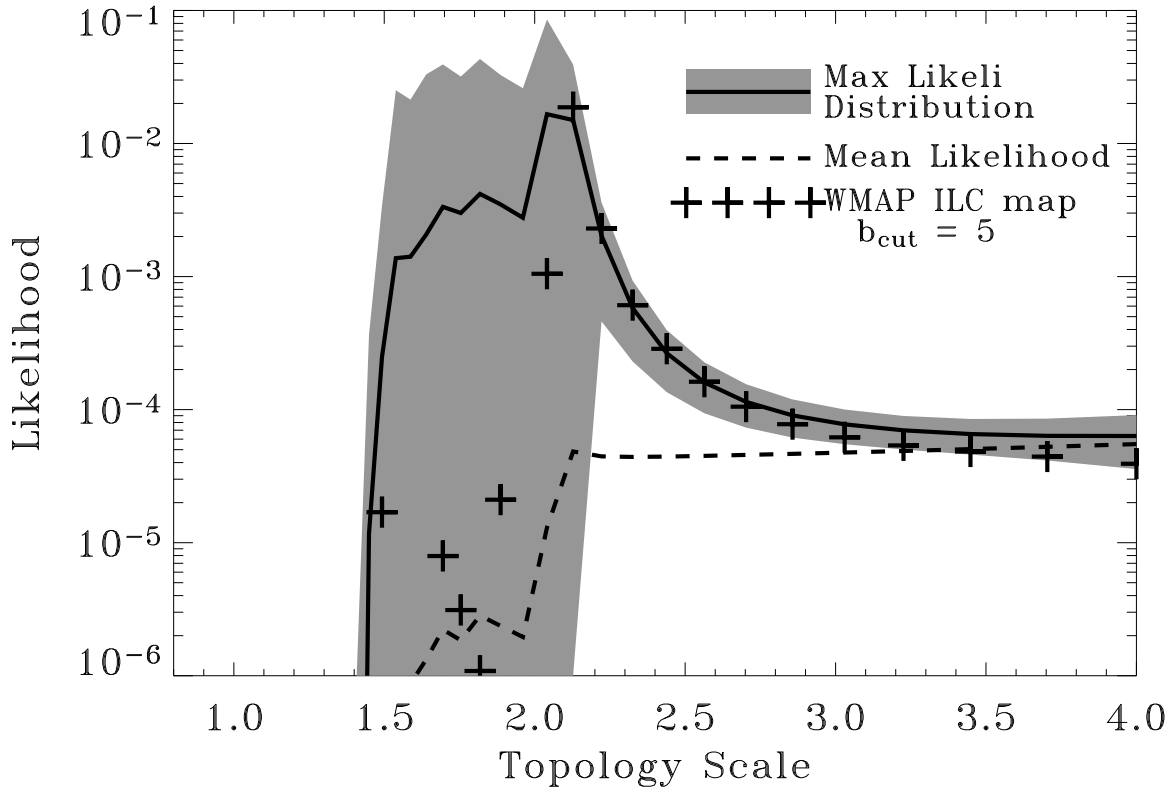}
\caption{Likelihood of the WMAP first-year data 
as a function of topology scale. 
At each topology scale $L$, 
the likelihood is maximized over the possible orientations 
of the fundamental domain.
For $L \sim 2$, chance alignments create a bias
in the likelihood estimator.
The solid line and grey band show the mean and standard deviation
of 1000 simulations drawn from an infinite flat model.
The WMAP data are consistent with the infinite model.
The dashed line shows the mean likelihood,
marginalized over orientation.
For the largest  topology scale,
there is no difference between maximizing or marginalizing 
over orientation; 
the scale is too large for the topology to break the global isotropy. 
At smaller scales, orientation starts to matter 
and although a a parent population may lack global isotropy, 
any given realization may ``appear'' to break it. 
The resulting bias can be quantified using Monte Carlo simulations.
}
\label{fig3-maxprob}
\end{figure}

\begin{figure}
\plotone{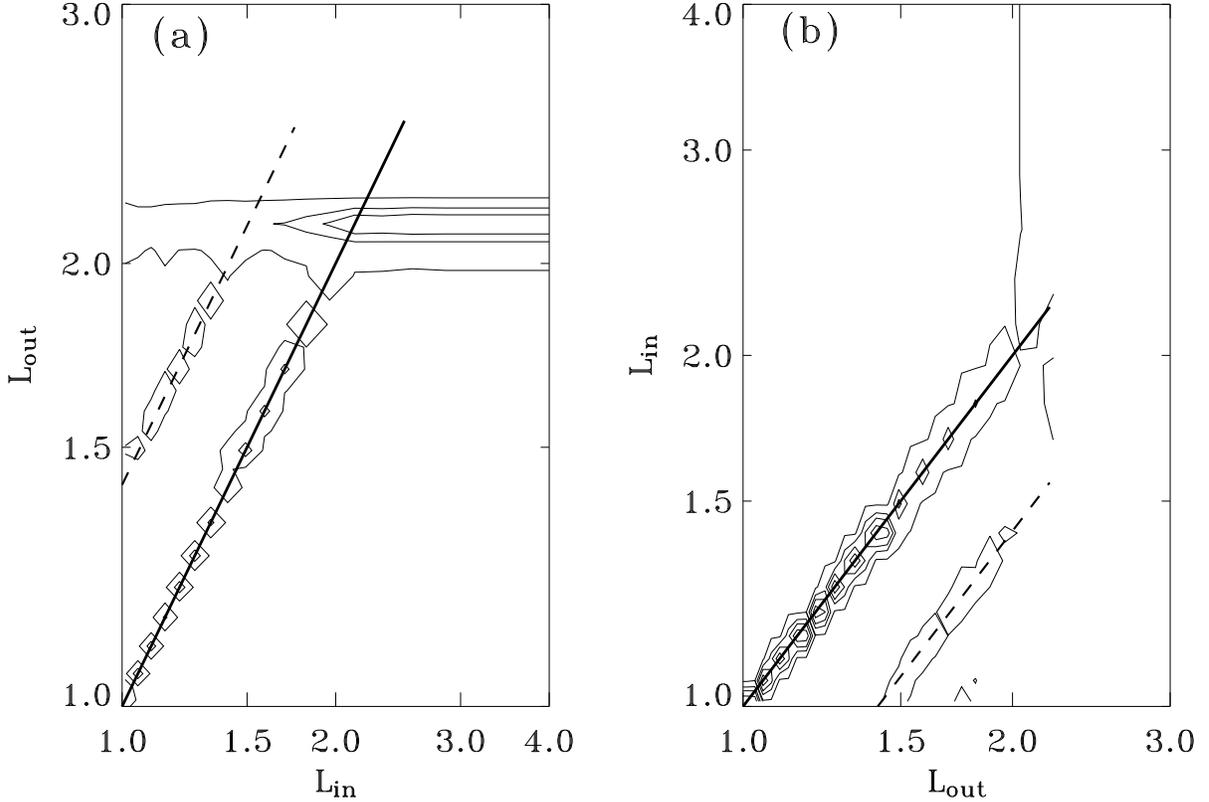}
\caption{Probability Distributions from Monte Carlo simulations. 
(a) Probability to obtain maximum-likelihood output $L_{out}$
as a function of the input scale size $L_{in}$ after maximizing
over orientation.
The solid line shows $L_{out} = L_{in}$.
For $L_{in} < 2$, the fundamental domain fits inside the
decoupling surface
and the distributions are strongly peaked at $L_{out} \simeq L_{in}$. 
For $L_{in} > 2$, maximizing over orientation
causes all the simulations to fall into a group 
centered around $L_{out} \approx 2.2$. 
(b) By inverting the relationship in (a), 
we obtain the probability that an observed output $L_{out}$
was drawn from a parent population with $L_{in}$.
Values $L_{out} > 1.96$
are indistinguishable from infinite models.
}
\label{fig4-probcontours}
\end{figure}

\begin{figure}
\plotone{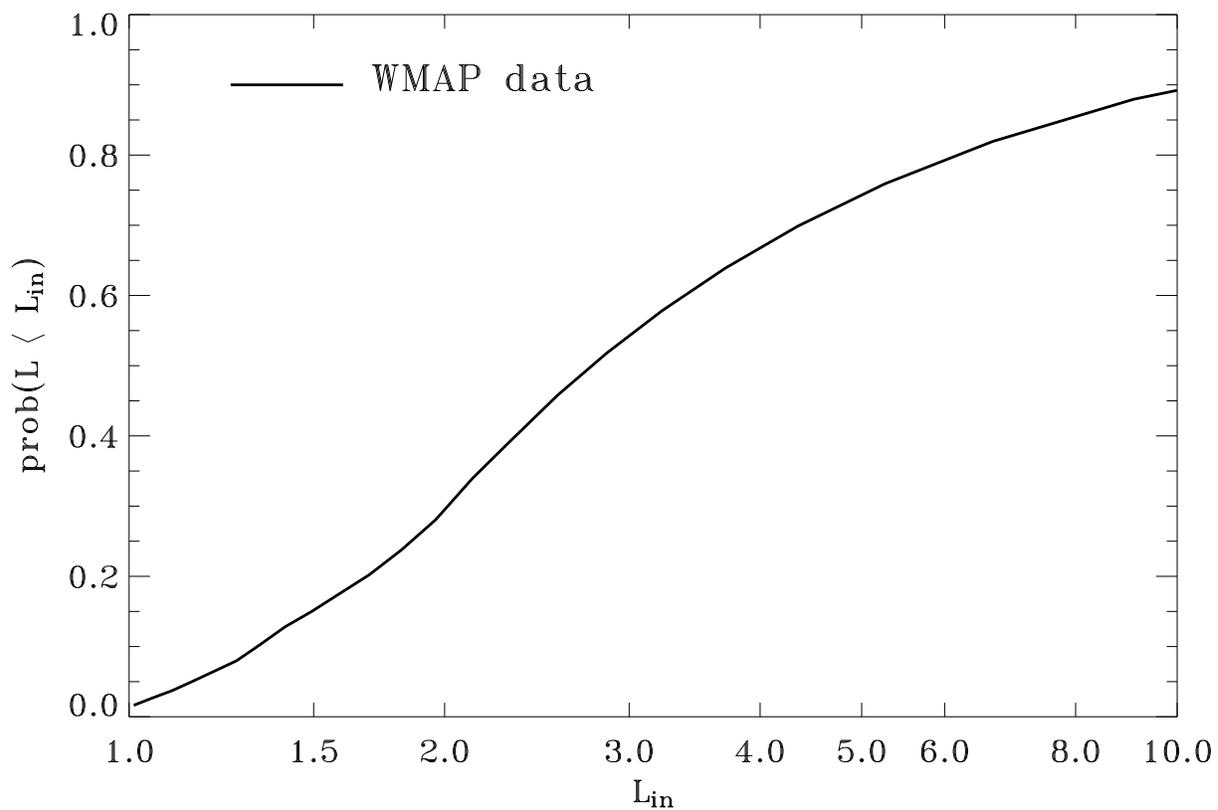}
\caption{Cumulative probability that the true topology scale
is below a given value.
WMAP data for galaxy cut $|b| > 5\arcdeg$.
We obtain limits $L > 2.1$ the distance to the decoupling surface
at 68\% confidence,
and $L > 1.2$ at 95\% confidence.
The WMAP data is consistent with a lack of any finite topology
for the cubic flat topologies we consider.
}
\label{fig5-wmapprob}
\end{figure}

The algorithm is sensitive both to the
power spectrum of the data
(diagonal elements of the covariance matrix for different ${a_{\ell m}}$)
as well as the phase information
contained in the off-diagonal elements.
For cell size $L < 2$
the off-diagonal elements are larger than the diagonal elements.
A comparison of the data to topological models
that utilizes only the power spectrum
can produce false positives
by ignoring the additional information
in the off-diagonal elements.
We demonstrate this using Monte Carlo simulations.
The power spectrum is rotationally invariant
and does not specify orientation.
We may thus modify Eq. \ref{chisq_def}
to use the
power spectrum $C_\ell$
and its covariance
(Eqs. \ref{power_spec_def} and \ref{eq-Cl-covar})
in place of the
spherical harmonic coefficients $a_{\ell m}$.
When only considering the power spectrum,
the maximum likelihood for the WMAP data
occurs at $L=1.1{\Delta\tau}$;
this is the ``finite'' model
power spectrum plotted in Fig. \ref{fig1-powerspectrum}.
Does this imply a positive detection of finite topology?
To test this, we generate 1000 Monte Carlo realizations 
drawn from a parent population with $L=1.1$ 
and generate the likelihood for each realization
using the full covariance matrix 
(Eqs. \ref{eq-alm-covar} and \ref{bcut_matrix_eq}).
For such a small topology scale, 
almost all realizations have their likelihood peak at $L=1.1$. 
This scale is small enough that the bias from maximizing over orientation 
is not important.
For each realization, 
we also generate a ``companion'' realization with exactly the
same power spectrum, 
but with completely uncorrelated ${a_{\ell m}}$'s.
The two realizations by construction
must give the same results 
for an analysis based solely on the power spectrum.
When we analyze the ``companion'' realizations 
using the full $a_{\ell m}$ covariance matrix,
we obtain results similar
to the infinite models displayed in Fig. \ref{fig3-maxprob}. 
A likelihood analysis using the full $a_{\ell m}$ covariance matrix
successfully distinguishes
models with compact topology
from models with identical power spectra
but without the correlations between different $a_{\ell m}$
required by the topology.
Suppression of power in the quadrupole and octopole moments
is a necessary but not sufficient condition
for a compact topology.
The WMAP data show reduced power at $\ell = 2$ and 3,
but do not show the correlations expected
for a compact topology
and are indistinguishable from infinite models.

For cell size comparable to the distance to the decoupling surface,
the correlations become weaker.
Maximizing the likelihood over orientation
then allows chance alignments
to introduce a bias
in the likelihood estimator.
We quantify this using Monte Carlo simulations.
The WMAP first-year data
are consistent with input models
drawn from parent populations with infinite fundamental domain.
We establish 95\% confidence limit
$L > 17$ Gps for the cell size
of a cubic topology, in agreement with the
result of 24 Gpc obtained by \cite{cornish/etal:2003}.

\acknowledgements
We thank G. Hinshaw for helpful discussions.
NGP thanks M. Doran for his help in using CMBEASY.
This work was supported by
the National Aeronautics and Space Administration
under the Astrophysics Data program
of the Office of Space Science.

\end{document}